\documentclass[aps,prl,superscriptaddress,twocolumn, showpacs,preprintnumbers,amsmath,amssymb]{revtex4-2}

\usepackage{hyperref}
\hypersetup{ colorlinks,
linkcolor=blue,
filecolor=blue,
urlcolor=blue,
citecolor=blue }

\usepackage{amsmath}
\usepackage{mathtools}
\usepackage{xcolor}
\usepackage{ulem}
\usepackage{xr}

\usepackage{subfigure}
\usepackage[pdftex]{graphicx}

\begin{document}

\title{Theoretical study of the competition between folding and contact interactions on the properties of polymers using self-avoid random walk algorithm}

\author{R. J. Santos Neto, A. A. Costa, P. F. Gomes}
\affiliation{Applied Complex Network Group of Jata\'i, Federal University of Jata\'i, BR 364 km 195, Jata\'i-GO, 75801-615, Brazil}

\begin{abstract}
The self-avoid random walk algorithm has been extensively used in the study of polymers. In this work we study the basic properties of the trajectories generated with this algorithm when two interactions are added to it: contact and folding interaction. These interactions represent the internal forces of the polymer as well as the effect of the solvent. When independently added to the algorithm, the contact interaction creates the compact phase while the folding one creates the extended phase. These are the consequences of the typical event of each interaction. On the other hand, when this typical event is avoided there is no established phase on the system. When simultaneously added, there is a competition between the interactions and the folding one is dominant over the contact one. The resulting phase is always the extended one with and without the contact interaction.

\end{abstract}


\maketitle

\section{Introduction}

Polymers are a type of long molecules with large applications in the industry and in the study of life, as proteins and other important molecules for life are polymers. Due to this essential role on modern societies and life in general, polymers have been subject of intense research for a long time \cite{Flory1949,Flory1953,Canevarolo2002,Gedde2019,Micheletti2011}. Generally speaking, the spatial conformation of the polymer chains defines its properties, which in turn defines its applications. On the other hand, the spatial conformation is defined by temperature and by the solvent. So it is important to understand how these external factors influence the spatial conformation of the polymers.

The phase diagram of a polymer has two main phases: extended (coil) and collapsed (globule or compact) phases. The former happens at high temperatures while the latter happens at low temperature and the transition is called $\Theta$-temperature \cite{Gennes1979,Duplantier1987,Grassberger1995,Grassberger1997,Oliveira2016}. Different theoretical methods have been used to study  the properties of polymer \cite{Rubin1965,Kholodenko1984,Maes1990,Vanderzande1998,Foster2001,Micheletti2021}. Many of these methods fall on the area of Statistical Mechanics, when probabilistic methods are evaluated and the results are the average values \cite{Fernandes2016,Vilela2020}. In this approach, one simple algorithm that has been extensively used to generate the trajectories representing the polymers is the Self-Avoid Random Walk, or SAW \cite{Chowdhury1985,Rockenbach2010}. This method has proven to be useful for study the phase diagram and the critical exponents \cite{Jensen2004,Clisby2010,Krawczyk2010}. It is a random walk with the restriction that any site can by occupied only once. The goal is to account for the excluded volume effect \cite{Madras2013}, which means that two monomers cannot occupy the same node on the lattice. Indeed, the original motivation for this type of walk was the study of polymers \cite{Montroll1950}.

Interactions are added to this algorithm to mimic the effect of the internal forces and the solvent. The most studied one is the contact interaction and its variations \cite{Narasimhan2001,Krawczyk2009,Rockenbach2010}. With this addition, the model remains probabilistic, however, the set of probabilities for the available monomers are not uniform anymore: one available monomer can be favoured or avoided. The interaction is inserted in the model using the Boltzmann factor $\beta = 1/kT$ \cite{Krawczyk2010}, so the effects compete with the temperature $T$. A second type of interaction is the folding one, which deals with the tendency of having straight lines on the chain \cite{Bastolla1997,Rockenbach2014}. When two interactions are added simultaneously on the model, the competition between them will define the phase of the system. 

Although this interaction has already been studied before, its basic features, temperature dependence and the competition with the contact interaction have not been described in a more didactical way. In this work we describe the general features of the folding and contact interactions itself and the competition between them. We analyze the influence of the limiting case of the temperature and the interaction control parameters. We observe that at low temperature the former creates an extended phase and the latter a compact one. When both are present, the folding interaction is dominant. In the next section we describe the algorithms we used to generate and the features to characterize the trajectories. In the following section we present our results and discussion. At last we present our conclusions.

\section{Methods}

The main part of the algorithm we use is the generator of a Self Avoiding Random Walk (SAW) in the 2d $(x,y)$ plane. In our model, a walk is a sequence of $n$ nodes (or monomers) in a square grid connected by edges. The allowed positions of the nodes are $(x,y)$ with both been integer numbers. The walk always begin at the center $(0,0)$. The second node can be any of the four nearest neighbors located at $(\pm 1,0)$ and $(0,\pm 1)$.  

\begin{figure}
    \centering
    \includegraphics[scale = 0.9]{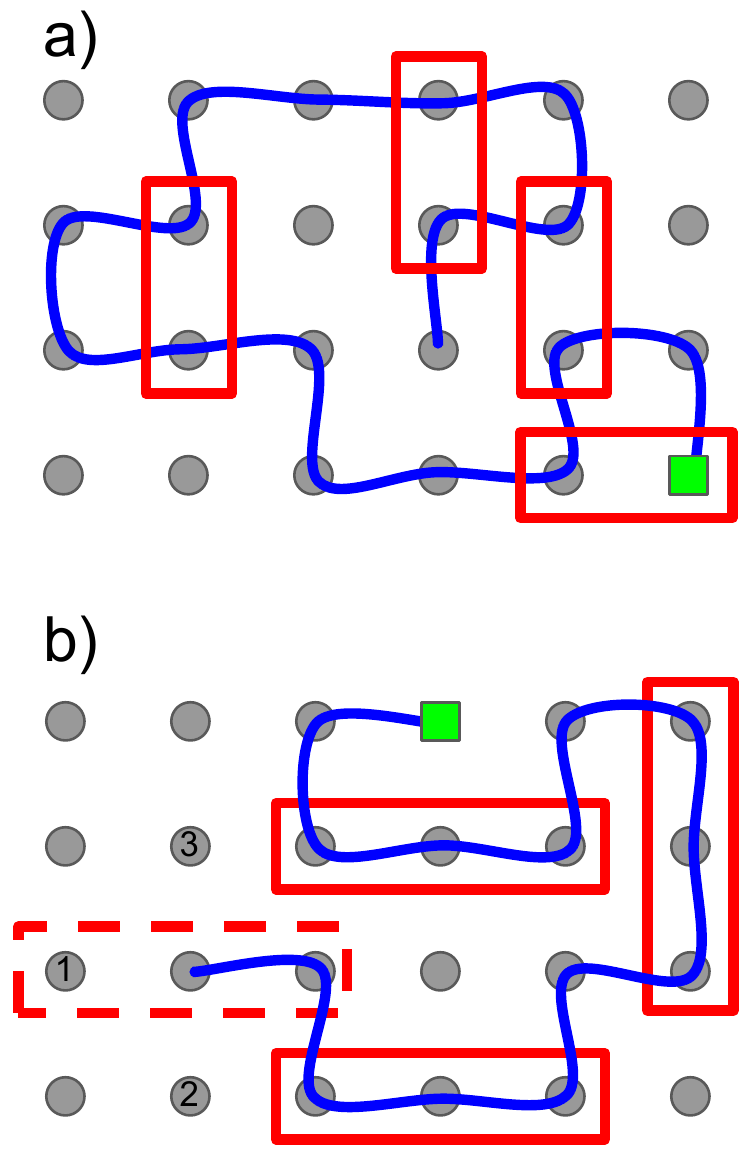}
    \caption{Illustration of trajectories (blue curve) on the square grid (gray circles) for the two interactions. The initial node is indicated by the green square. The typical events are indicated by the red rectangles. (a) Trapped trajectory with four contacts. (b) Trajectory with three straight lines.}
    \label{contatos_retas_v5}
\end{figure}

The self-avoiding part of the walk refers to the fact that each node is visited only once by the walk \cite{Madras2013}. So crosses of the trajectories are not allowed. In this way, from the third node of the chain, there are at maximum three available nodes. The walker never goes back to the node it came from. This creates a problem that the chain can be trapped in itself when there is no available nodes to be added to the chain. The control parameter $N$ is the desired number of nodes, it is reached when there is no trapping: $n=N$. However if the chain gets trapped in any moment, then we have $n<N$. Figure \ref{contatos_retas_v5}(a) indicates such a case. At each step in the construction of the SAW, one node is added to the chain. In the SAW with no interaction the available nodes in each step have equal probability to be selected.

\subsection{Interaction: typical event}

The effect of the interaction is to change the probabilities of the available sites to be added to the chain. The interactions are added to the SAW using the Boltzmann factor $e^{-\beta E}$ where $E$ is the energy of the interaction and $\beta = 1/(kT)$. The temperature is $T$ and the Boltzmann constant is $k$. In high temperatures (low $\beta$) the interactions become irrelevant while on low temperatures (high $\beta$) the interactions are dominant.

Each interaction has a typical event (will be defined later), which can be induced or avoided by the interaction, depending on the sign of the energy $E$. The probability for the available site $j$ been selected is given by:
\begin{equation}
p_i = \frac{e^{-\beta E_i}}{\sum_{j=1}^{\mu} e^{-\beta E_j}}. \label{probability}
\end{equation}
The denominator, similar to a partition function, ensures that the total probability is 1.0. As we are considering a square lattice, the amount of available nodes is $\mu = 1,2$ or 3. The trapping case corresponds to $\mu = 0$. The parameter $E_i$ gives the information if the node $i$ creates a typical event or not:
\begin{equation}
E_i = \left\{
\begin{array}{rl}
 E_0 \neq 0, &  \text{ typical event}  \\
0, &  \text{ otherwise.}
\end{array} 
 \right.  \nonumber 
\end{equation}
If we choose $E_0=0$, it is the no interaction case. Once the probabilities are calculated, the next node is chosen accordingly. 

The first interaction is the contact one and the typical event is the contact. Its definition is: two nearest neighbors nodes but not consecutively bound on the chain (or trajectory). Figure \ref{contatos_retas_v5}(a) illustrates this event. The second interaction is the folding one and it is related to the possibility that three consecutive sites in the chain make a straight line or make a folding. It is a matter of choice to define the typical event as the straight line or as the folding. Following the tradition, let's define the event as the straight line. Figure \ref{contatos_retas_v5}(b) shows an illustration of this typical event, where we can see that the last node has 3 available nodes to connect: nodes labeled 1, 2 and 3. If, and only if, the node 1 is selected another straight line will be formed.

Four consecutive sites making a straight line constitutes two events.  Equation \ref{probability} is used for any interaction, the difference is the energy $E$ which assumes a given value for each typical event. For simplicity, we use $E_i=c_i \varepsilon$ for the contact interaction where $c_i$ is the number of contacts of node $i$. In the same way, $E_i=\ell_i \chi$ for the folding one with $\ell_i$ been the number of straight lines of node $i$. When the two interactions are added simultaneously, the probability is still given by Eq. \ref{probability} with $E_i = c_i \varepsilon + \ell_i \chi$.


\subsection{Order parameters}

To identify the effect of the interactions on the polymers we use 4 different features. The first one is the number of contacts on the chain, motivated by the typical event of the contact interaction. The second one is the number of 3-node straight lines, again, motivated by the folding interaction. In order to be able to compare this quantities, we use the normalized version of each of them, dividing by the total number of nodes $N$ of the chain. So the normalized number of contacts (or density) is $c$ while the normalized number of straight lines is $\ell$. In this way each quantity has a maximum of 1.0, besides the minimum of 0.0.

The others parameters we use are the success rate $R$ and the anisotropy $\delta$. The first one is defined as the number of successful chains divided by the total number of generated samples $S$. A successful chain is the one with $n=N$, which means, the final number of nodes is the desired one, no trapping happened. For the second, let $n_x$ and $n_y$ be the number of horizontal and vertical edges on the chain. An edge is the straight line connecting two nodes. The anisotropy is defined as \cite{Krawczyk2010}:
\begin{equation}
    \delta = 1.0 - \frac{\min (n_x,n_y)}{\max (n_x,n_y)},
\end{equation}
where $\min (a,b)$ and $\max (a,b)$ are the minimum and maximum, respectively, of the numbers $a$ and $b$. This anisotropy measures the orientational order of the chains. For example, $n_x = 0$ means there is no horizontal edge, so the chain is completely vertical. This is a maximum in the orientational order. If $n_x=n_y$, we have a straight line making 45 degrees with the $x-$axis, also a maximum in the orientational. On the other hand if $n_x$ and $n_y$ have no connection between them on the averagem, we have no orientational order on the system. In this case we have on the average a $\delta$ close to zero as the system increases. If this limiting value as the system ($N$) increases is non zero and less than 1.0, we have a weak orientational order. If this limiting value is 1.0, it is an evidence of strong orientational order.

\begin{figure*}
    \centering
    \includegraphics[scale = 0.7]{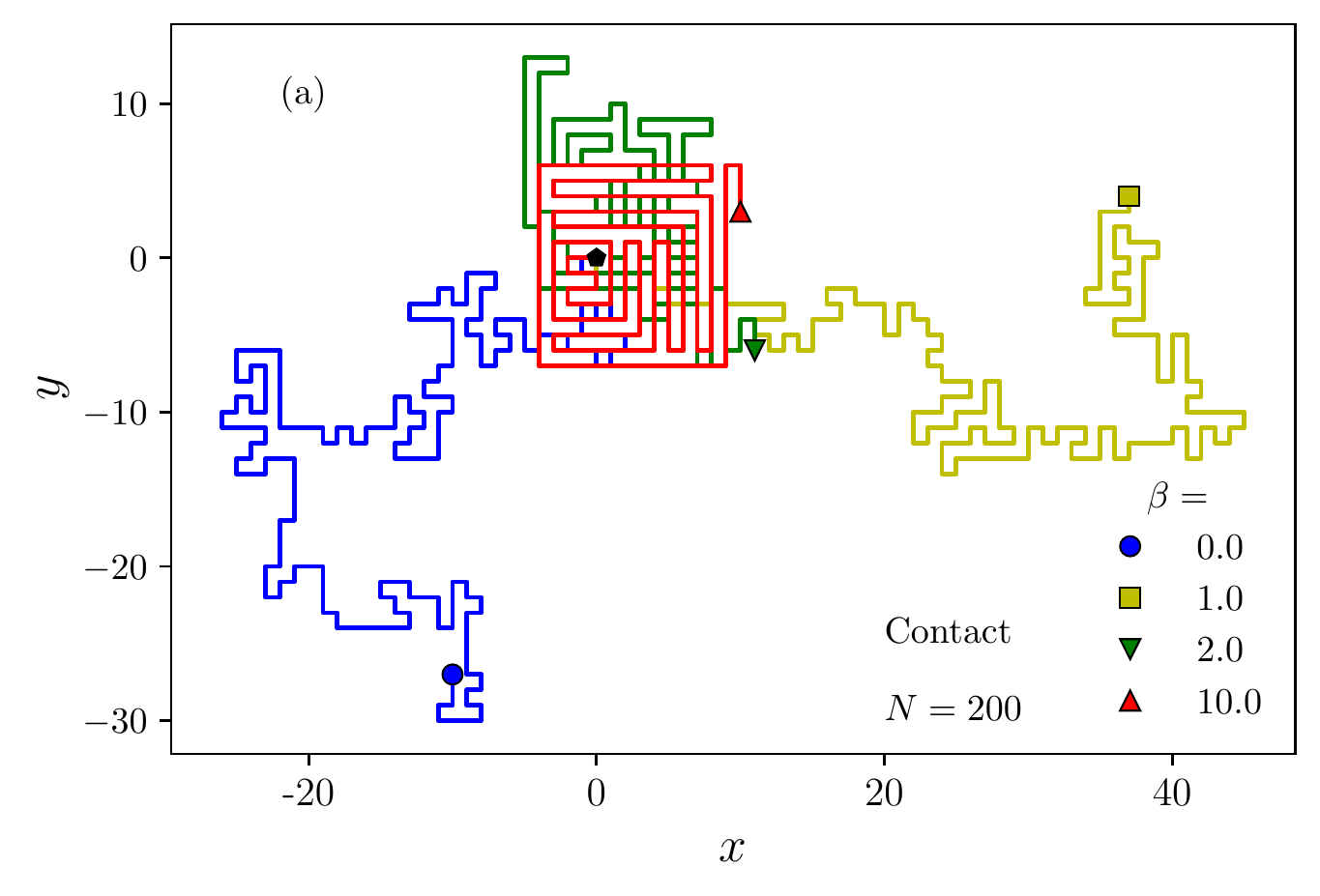}
    \includegraphics[scale = 0.7]{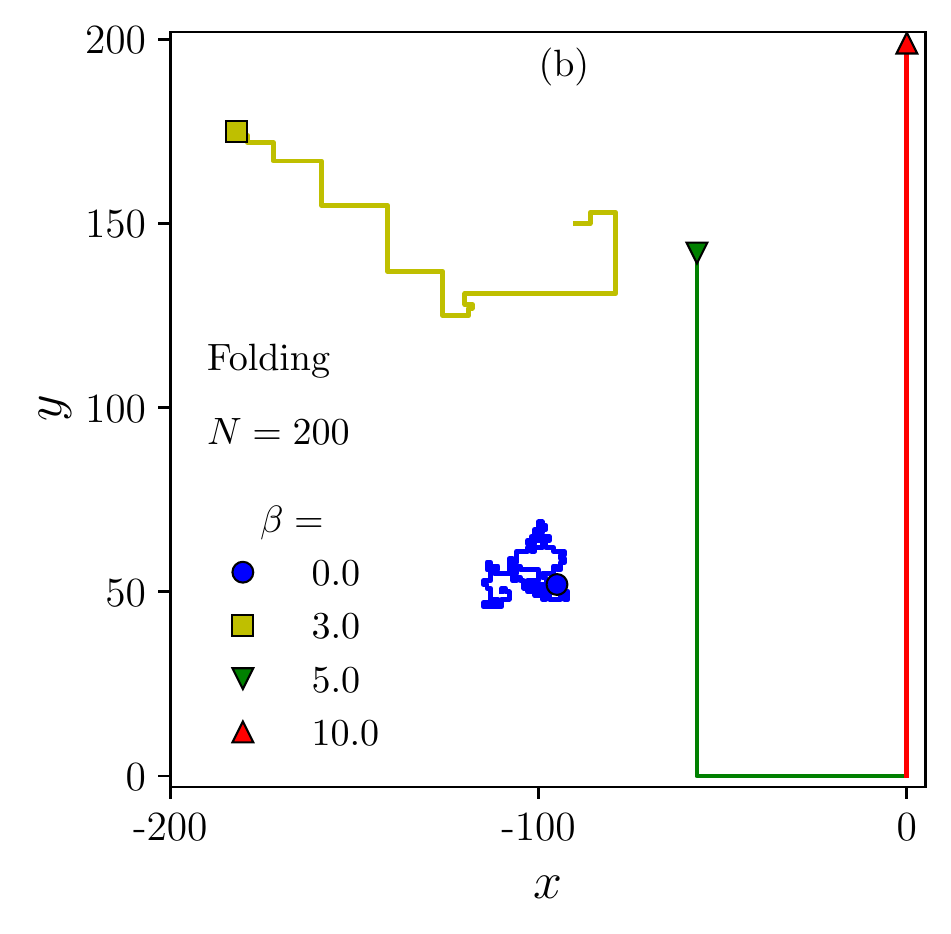}
    \caption{Typical trajectories for each interaction for different values of $\beta$. The initial node is at the origin and the marker indicates the final position. Length of the chain: $N=200$. (a) Contact interaction: $\varepsilon = -1.0$ and $\chi = 0.0$. (b) Folding interaction. $\varepsilon = 0.0$ and $\chi = -1.0$ Trajectories has been shifted for better visualization.}
    \label{figura2}
\end{figure*}

\section{Probabilities}

Let's dig in more detail on the calculation of the probabilities. Suppose we have in a given moment of the walk 3 available nodes for the next step. One of these, say node 1, makes a typical event of the interaction (contact or folding): $c_1$ or $\ell_1$ equal to 1.0 and all other equal to zero. So we have: $E_1  = E_0$ and $E_2= E_3= 0$. The denominator of Eq. \ref{probability} is:
\begin{equation}
\sum_{j=1}^3 e^{-\beta E_j} = 2+e^{-\beta E_0}. \nonumber
\end{equation}
The probabilities are: 
\begin{eqnarray}
    p_1 &=& \frac{e^{-\beta E_0}}{2+e^{-\beta E_0}}, \nonumber \\
    p_2 &=& p_3 = \frac{e^{-\beta E_2}}{2+e^{-\beta E_0}} = \frac{1}{2+e^{-\beta E_1}}, \nonumber 
\end{eqnarray}
Let's analyze this probabilities as function of $\beta$ e $E_0$.

\subsection{Case $\beta \approx 0$}

If the temperature is very high ($T \rightarrow \infty$), $\beta$ is very low: $\beta \rightarrow 0$. In this limit: $e^{-\beta E_0} \sim e^0 = 1$. So, the probabilities become:
\begin{eqnarray}
\lim_{\beta \rightarrow 0} p_1 &=& \frac{ e^0}{2+e^0} = \frac{1}{3}, \nonumber \\
\lim_{\beta \rightarrow 0} p_2 &=& \frac{ 1}{2+e^0} = \frac{1}{3}. \nonumber\end{eqnarray}
Thus $p_1=p_2=p_3$. At high temperatures, the interactions do not make any difference at all.

\begin{figure}
\centering
\includegraphics[scale = 0.5]{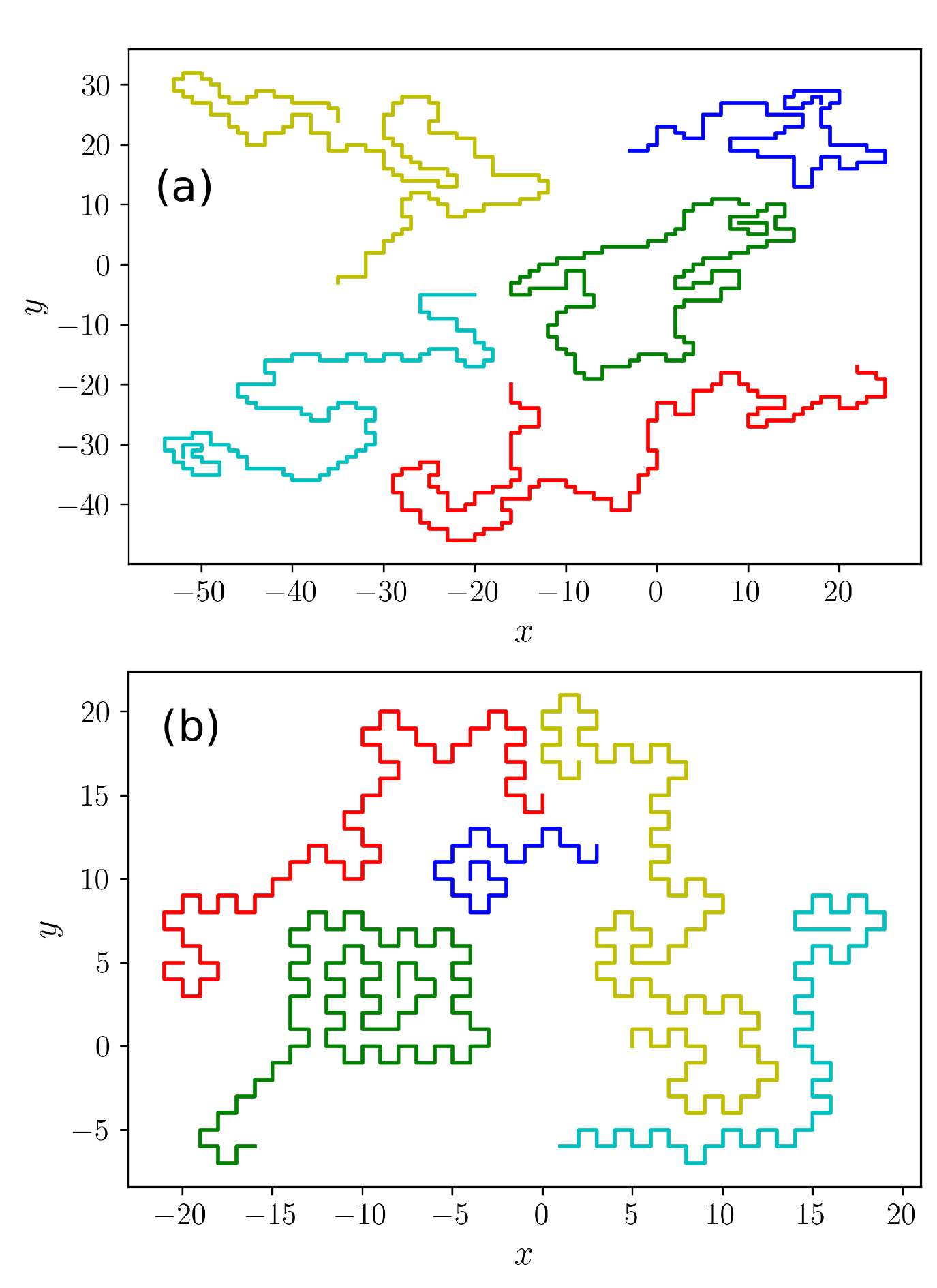}
\caption{Typical trajectories avoiding the typical event of each interaction. Trajectory are shifted for better visualization. Parameters: $N=200$ and $\beta = 10$. (a) Contact interaction: $\chi = 0.0$ and $\varepsilon=1.0$. (b) Folding interaction: $\chi = 1.0$ and $\varepsilon=0.0$.}
\label{figura3}
\end{figure}

\begin{figure*}
    \centering
    \includegraphics[scale = 0.42]{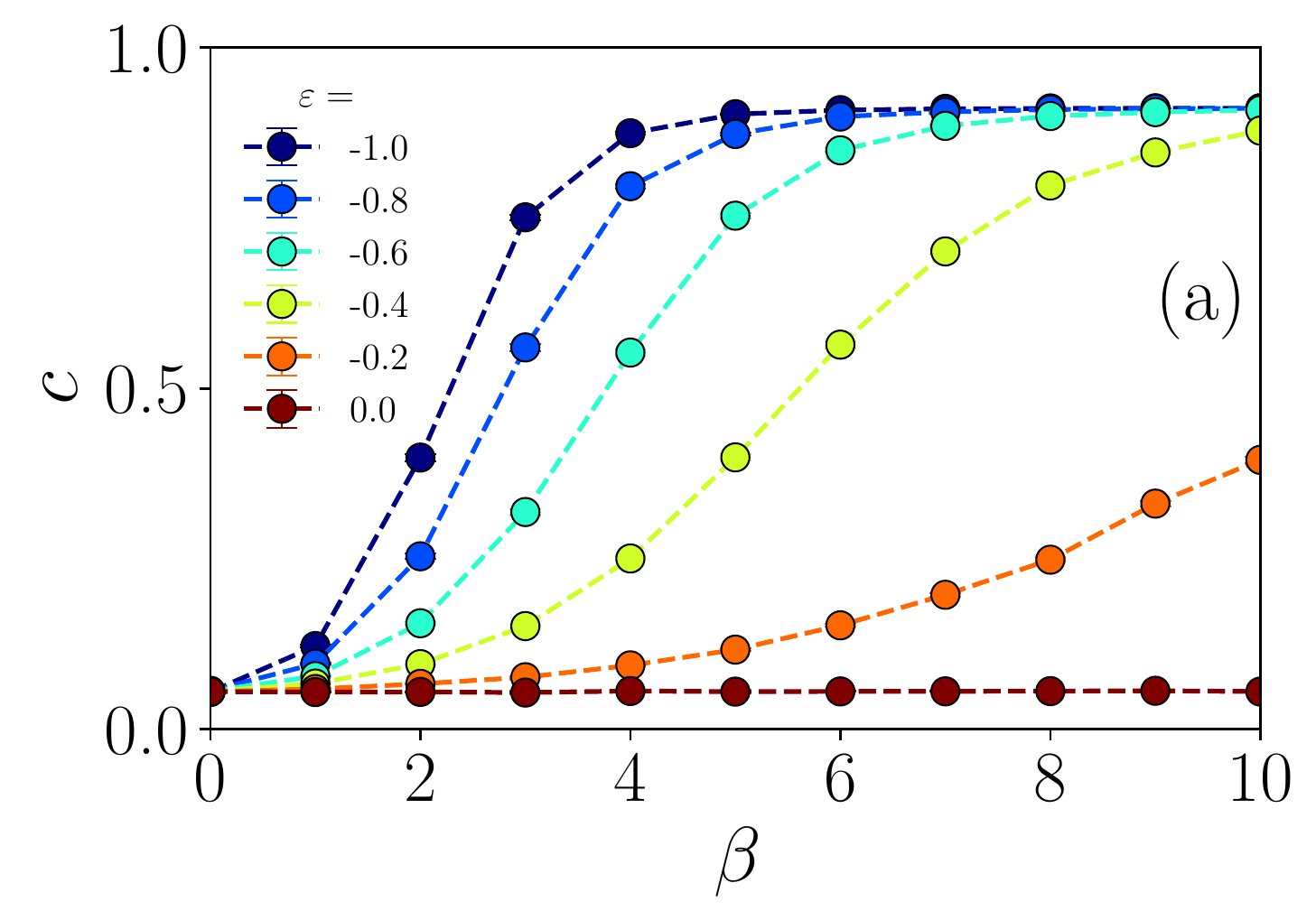}
    \includegraphics[scale = 0.42]{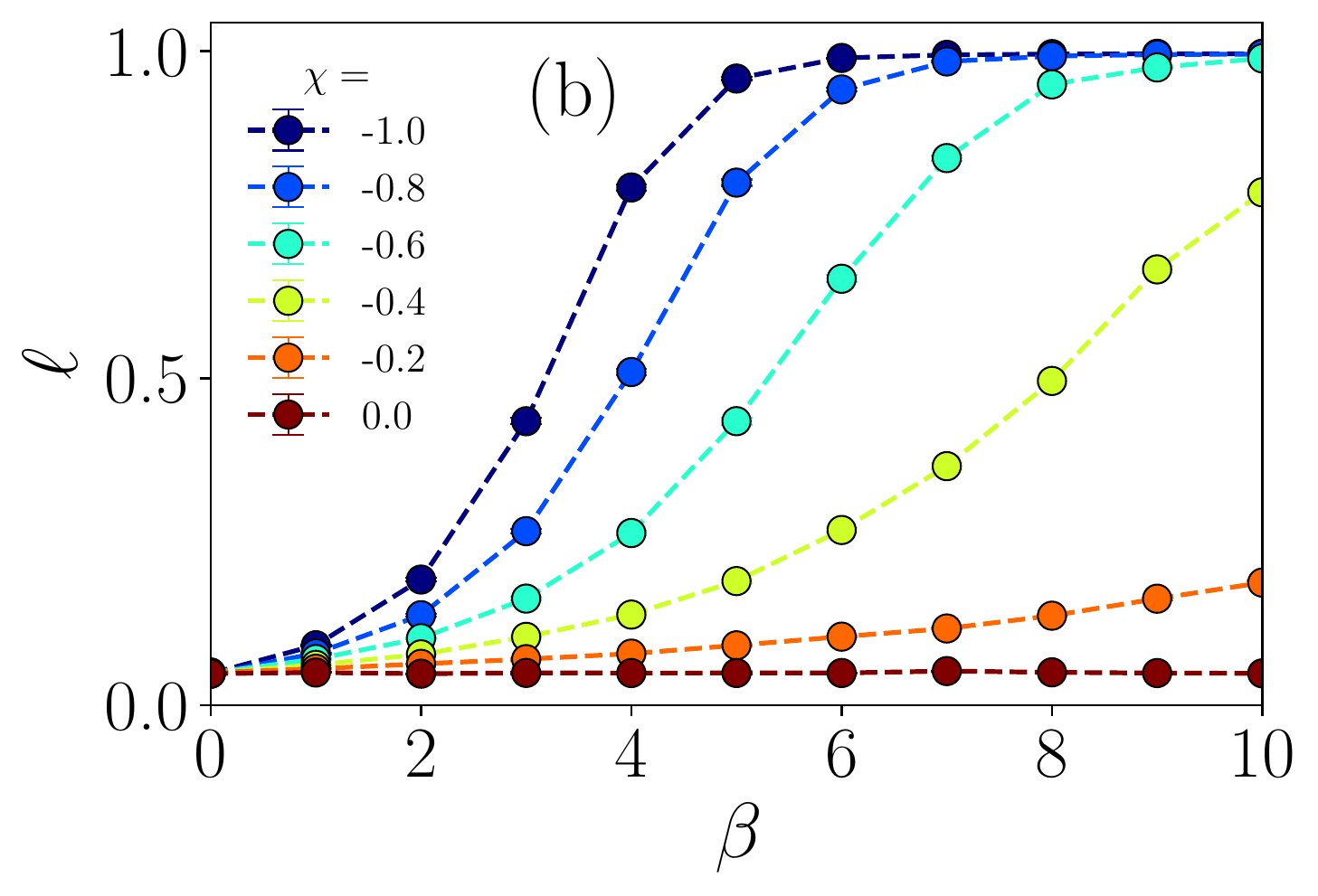}
    \includegraphics[scale = 0.42]{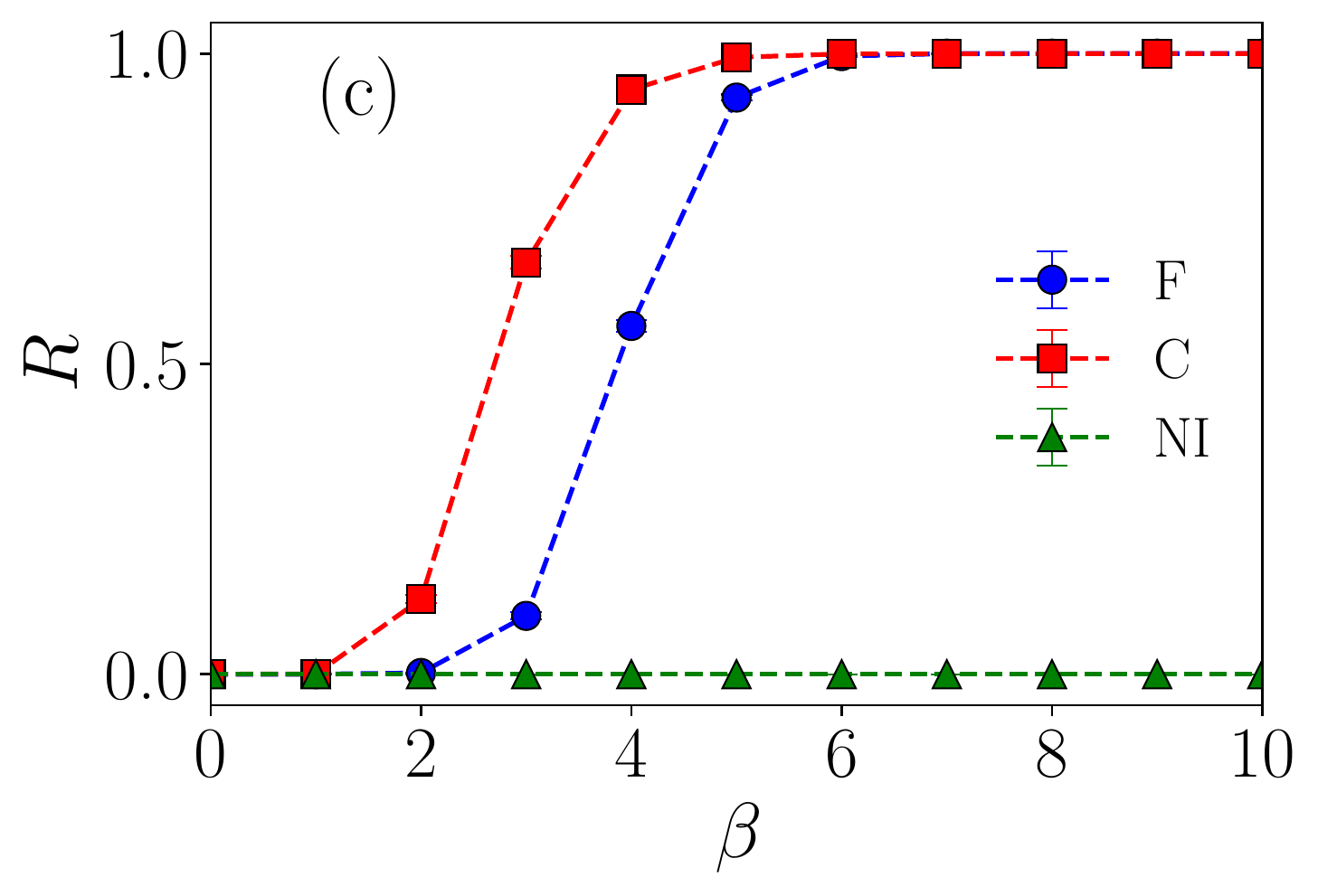}
    \includegraphics[scale = 0.42]{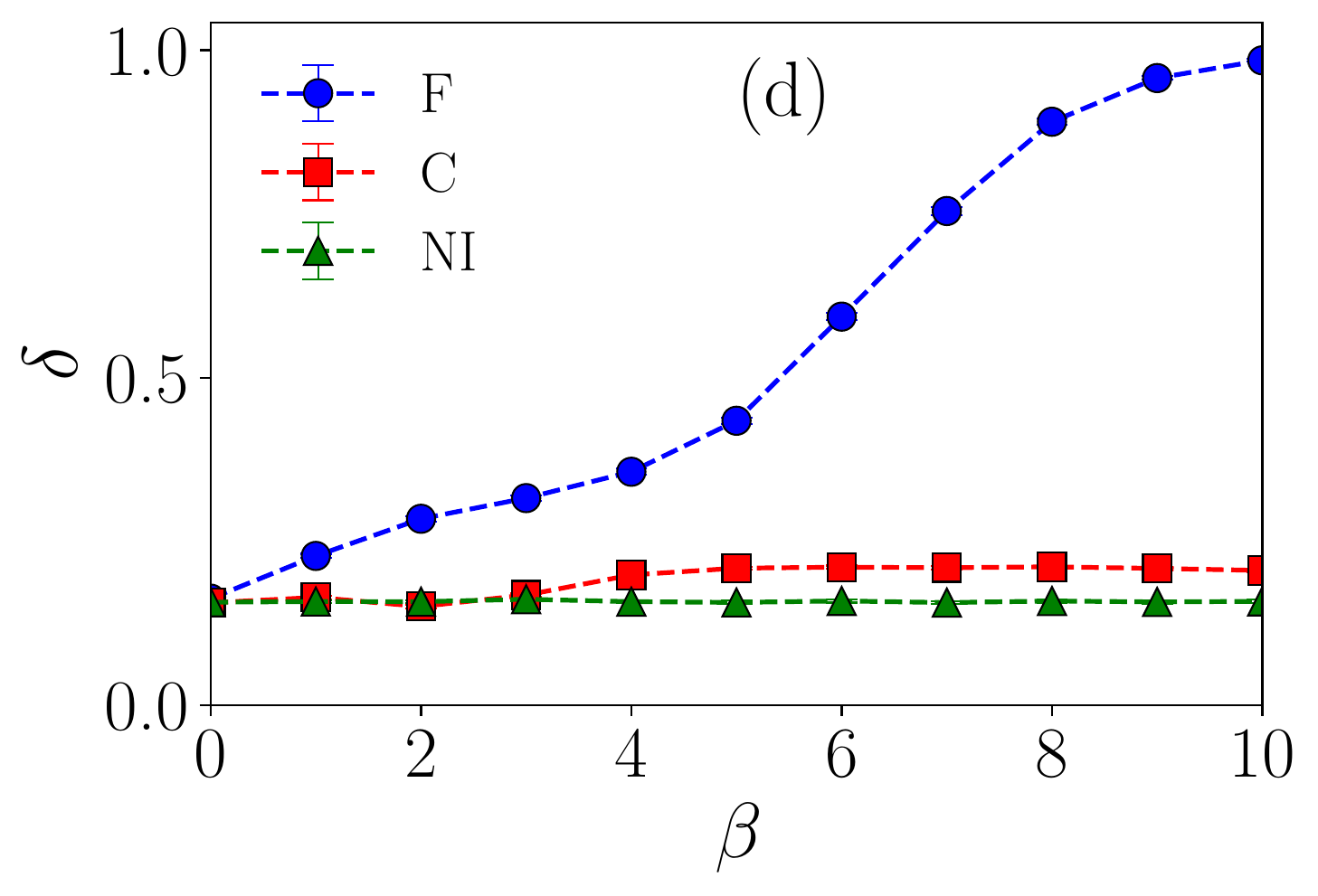}
    \caption{Typical order parameters of each interaction. Parameters: $N=500$, $S_1 = 25$ and $S_2=100$. The lines connecting the points are just a guide to the eyes. (a) Number of contacts $c$ for different interaction intensity $\varepsilon$. (b) Number of straight lines $\ell$ for different interaction intensity $\chi$. (c) Success rate $R$ and (d) anisotropy $\delta$ for the three types of interaction: F $=$ folding interaction ($\varepsilon = 0.0$ and $\chi = -1.0$), C $=$ contact interaction ($\varepsilon = -1.0$ and $\chi = 0.0$) and NI $=$ no interaction ($\varepsilon = \chi = 0.0$).}
    \label{figura4}
\end{figure*}

\subsection{Case $\beta >> 1$ and $E_0 > 0$}

The interactions will matter only when temperature is low, which means large $\beta$. Let's see what happens when $E_0>0$ in this limit. With these two conditions the product $\beta E_0$ is positive, so the exponential $e^{-\beta E_0} = \eta$ will be small. The probabilities become:
\begin{eqnarray}
\lim_{E_0>0,\beta \rightarrow \infty} p_1 &=& \lim_{\eta \rightarrow 0} \frac{\eta}{2+\eta} \sim 0, \nonumber \\
\lim_{E_0>0,\beta \rightarrow \infty} p_2 &=& \lim_{\eta \rightarrow 0} \frac{1}{2+\eta} \sim \frac{1}{2}. \nonumber
\end{eqnarray}
Thus: $p_1=0$ and $p_2=p_3=1/2$, which means that the typical event (node 1) is avoided in this limit. All the other nodes have the same probability to be chosen.

\subsection{Case $\beta >> 1$ and $E_0 < 0$}

Now $\beta E_0 <0$ and the exponential $e^{-\beta E_0} = \eta$ becomes too large:
\begin{eqnarray}
\lim_{E_0<0,\beta \rightarrow \infty} p_1 &=& \lim_{\eta \rightarrow \infty} \frac{\eta}{2+\eta} \sim 1.0, \nonumber \\
\lim_{E_0<0,\beta \rightarrow \infty} p_2 &=& \lim_{\eta \rightarrow \infty} \frac{1}{2+\eta} \sim 0.0. \nonumber
\end{eqnarray}
Thus the typical event will happen for sure in this limit ($p_1=1.0$) while the others nodes are avoided ($p_2=p_3=0.0$).

\section{Results}

We have used a simple algorithm to generate the trajectories, sometimes called true or myopic algorithm. Each trajectory is a group of $n$ pairs $(x,y)$ and the distance between two consecutive points is always 1. This trajectory can be obtained considering a traveler randomly moving from one point to another. In the general method, the traveler avoids a given position that he already visited \cite{Amit1983}. In this work we consider that the traveler never moves back to a position he already visited, which by itself, is some times called self-repelling \cite{Grassberger2017}. 

For each interaction, the typical event can be achieved with a large $\beta$ or with a large negative value of the control energy $E_0$ ($\varepsilon$ for the contact and $\chi$ for the folding interaction). So, we define some boundaries for this control energy for simplicity: $-1.0 < \varepsilon,\chi < 1.0$. The limit $-1.0$ refers to the case with the strongest intensity of the interaction: most probable to appear the typical event. The opposite limit $+1.0$ also refers to the strongest intensity in order to avoid the typical event. The case $0.0$ refers to no interaction. Both interactions are independently added to the algorithm.

The data, the analysis and the graphics shown here were developed using Python with the packages: Numpy \cite{Harris2020}, Matplotlib \cite{Hunter2007} and Pandas~\cite{McKinney2010}. All codes and data are available upon request. 

\subsection{Typical trajectories}

As we have seen earlier $\beta = 0$ eliminates the effect of the interactions. When $\beta > 0$, the sign of the control parameter $E_0$ defines if the typical event will be favoured or avoided. In the case of contact interaction we use $E_0 =\varepsilon$ and for the folding interaction $E_0 = \chi$. We included both interactions in the algorithm, so the probability remains defined by Eq. \ref{probability} with $E_0 = \varepsilon c_i + \chi \ell_i$.

Figure \ref{figura2}(a) shows typical trajectories for contact interaction. For increasing $\beta$ the trajectories start to change, becoming more compact and for $\beta \geq 10$ there is no further significant changes. In the same way for the folding interaction (Fig. \ref{figura2}(b)) the effect is visible and again, there is no more significant changes for $\beta \geq 10$. However, for this interaction the trajectory become extended.

For the sake of completeness, we also show the typical trajectories avoiding the typical event for each interaction. This is achieved by setting the correspondent energy as $+1.0$. Figure \ref{figura3}(a) shows the case for the contact interaction, where it is visible the small quantity of contacts. The trajectories are not compact as before. On the case of folding interaction on Fig. \ref{figura3}(b) shows that the trajectories are not extended as before, and no straight line is visible.

\begin{figure*}
    \centering
    \includegraphics[scale = 0.94]{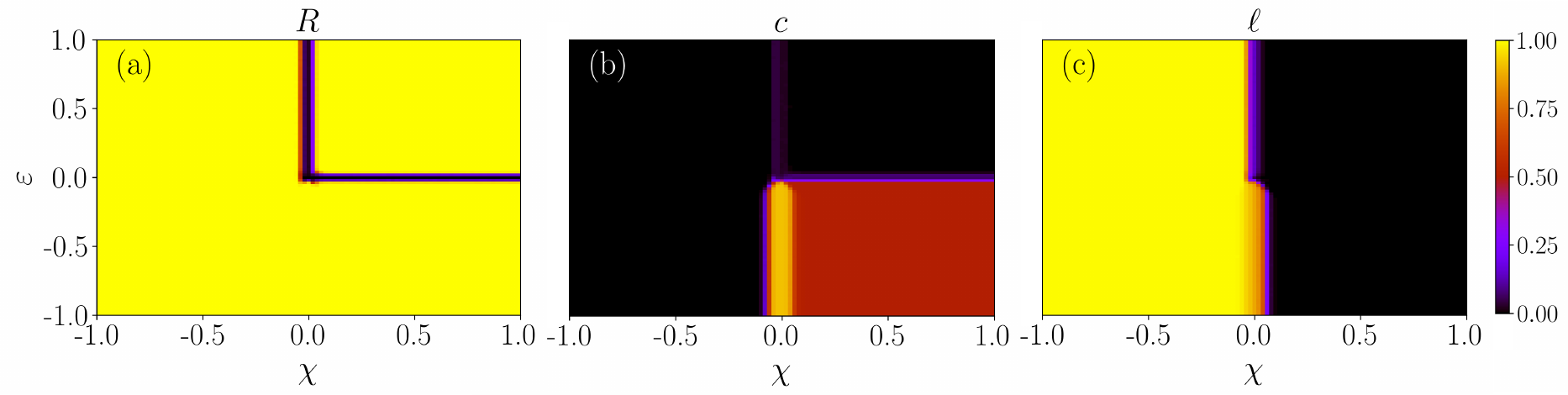}
    \caption{Contour plot of different order parameters on the $\chi,\varepsilon$ plane. Parameter: $N=500$, $\beta = 100$, $S_1 = 50$ and $S_2=400$. (a) Success rate $R$. (b) Number of contacts $c$. (c) Number of straight lines $\ell$.}
    \label{figura5}
\end{figure*}

\subsection{Order parameters}

Now we show how the features of the system change with the temperature between the two limits: $\beta = 0.0$ and $\beta >> 1.0$. First, Fig. \ref{figura4}(a) shows the typical order parameter of the contact interaction: the normalized number of contacts $c$. We can see that $c$ increases with the interaction $\varepsilon$ for large $\beta$. At $\beta \geq 6.0$ the parameter is at its maximum at large $\varepsilon$: $c \sim 0.9$. As this condition corresponds to the compact phase, we see that $c$ is maximum in this phase. The equivalent behavior is observed for the normalized number of straight lines $\ell$ with folding interaction at Fig. \ref{figura4}(b). Again, at $\beta \geq 6.0$ and large $\chi$ we have $\ell$ at its maximum around 1.0, corresponding to the extended phase. These two results show that indeed each order parameter is maximum in the corresponding phase.

The success rate $R$ can also be used as an indication for either interactions, as shown on Fig. \ref{figura4}(c). Again, at $\beta \geq 6.0$ we have $R$ at its maximum 1.0 for both interactions even though the symmetry in each case are different. The anisotropy $\delta$ highlights this difference in the symmetry, as shown on Fig. \ref{figura4}(d). In the case of folding interaction the extended trajectory has a larger anisotropy, in contrast to the compact trajectory of the contact interaction. However, it is still possible to distinguish between the contact interaction and no interaction case: the former has a slightly larger anisotropy than the former.

\subsection{Competition between the interactions}

When the two interactions act simultaneously on the system, there is a competition between them and the final configurations are different. In order to better study this competition we calculated the order parameters on the $(\chi,\varepsilon)$ plane at the region $-1.0<\chi,\varepsilon < +1.0$. We used $\beta = 100$ to enhance the effect of the interactions. Figure \ref{figura5}(a) shows the success rate $R$ on this region. We have $R \sim 0$ when $\chi$ or $\varepsilon$ is positive and the other is zero. Or, when only one typical event is avoided. $R$ reaches its maximum value of 1.0 on the rest of the space, even when the two typical events are avoided ($\varepsilon = \chi = +1.0$). 

The number of contacts $c$ is plotted in the same region at Fig. \ref{figura5}(b). It is different from zero only in the fourth quadrant: $\chi >0$ and $\varepsilon < 0$. From this we can say that the folding interaction is dominant over the contact one. So if $\chi < 0$ we have $c \sim 0.0$ regardless the value of $\varepsilon$. When $\chi >0$ the contact interaction has a fighting chance and some contacts are observed. The only region where $c \sim 1.0$ happens with $\chi \sim 0.0$ and $\varepsilon < 0.0$. The number of straight segments $\ell$ is plotted in the same region as shown at Fig. \ref{figura5}(c). As expected, we observe $\ell \sim 1.0$ in the entire region $\chi < 0.0$, independent of the value of $\varepsilon$, which reinforces the stronger character of the folding interaction over the contact one.

\begin{figure*}
    \centering
    \includegraphics[scale = 0.34]{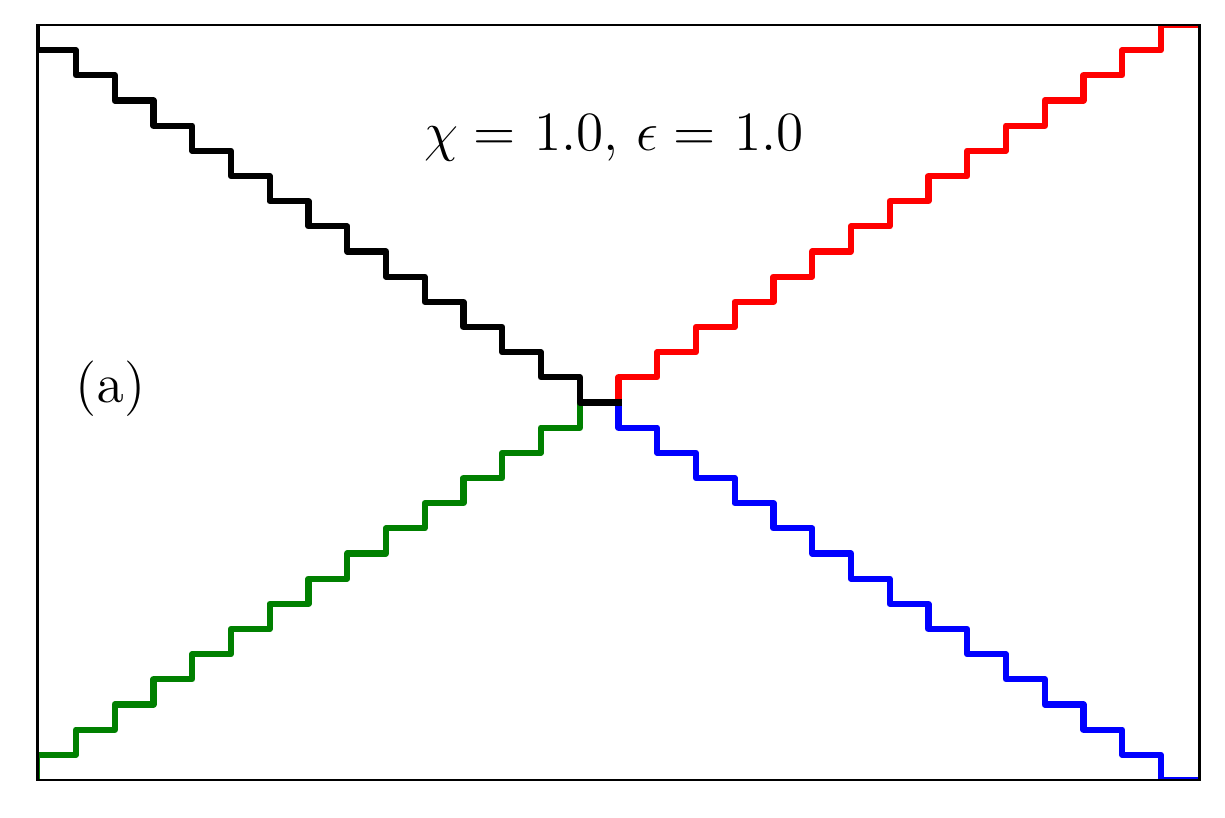}
    \includegraphics[scale = 0.34]{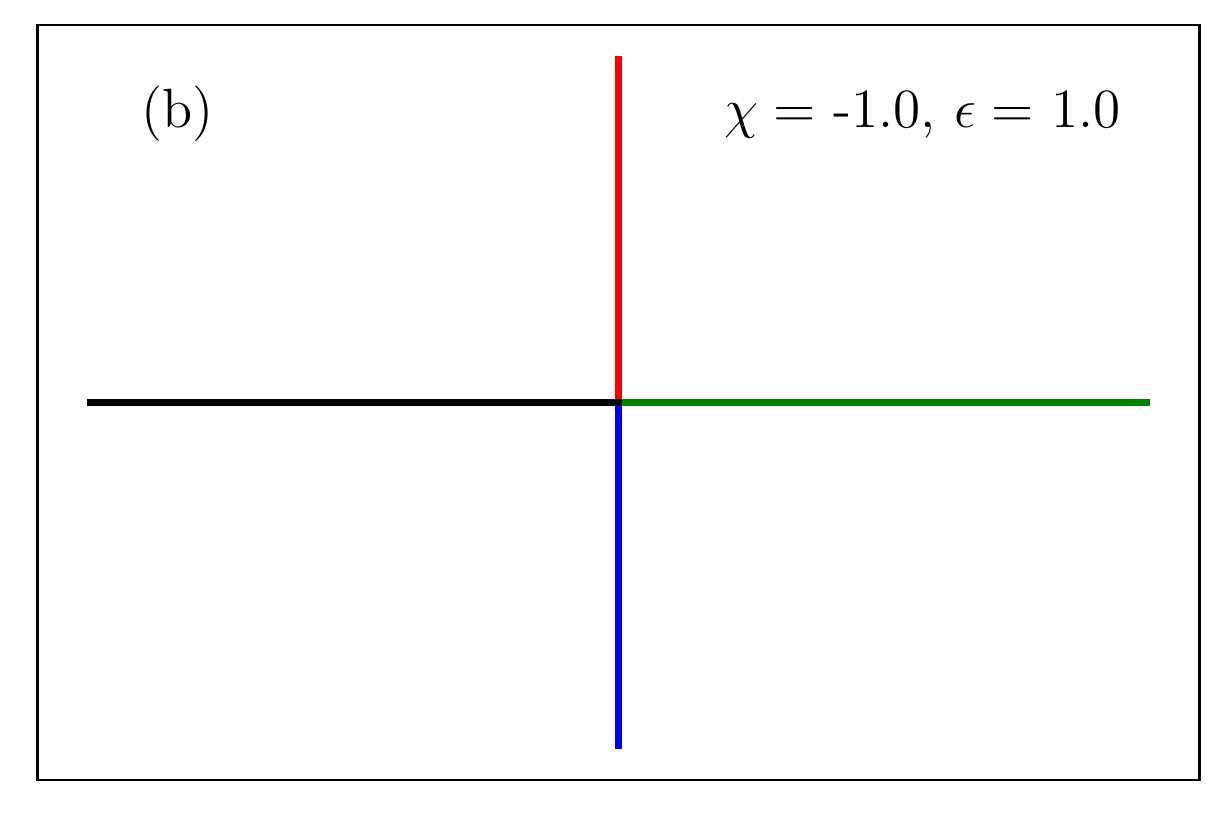}
    \includegraphics[scale = 0.34]{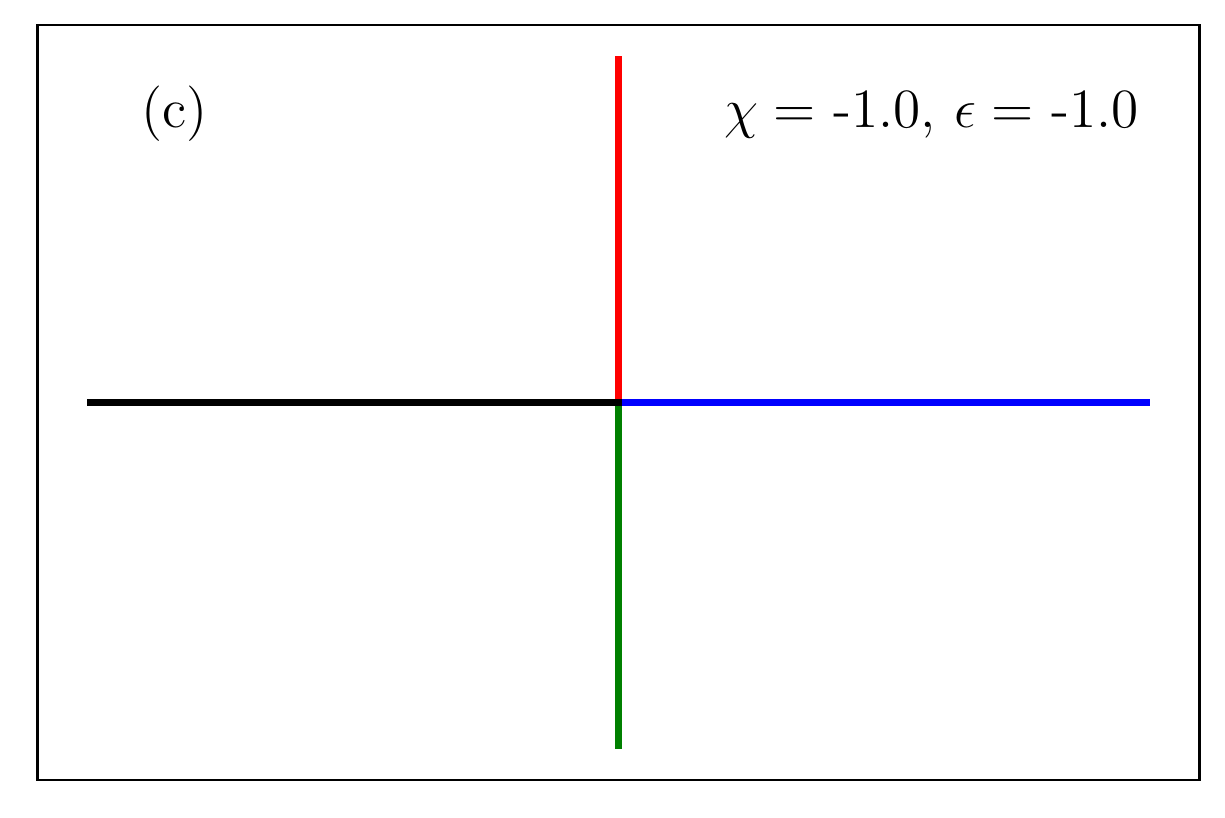}
    \includegraphics[scale = 0.34]{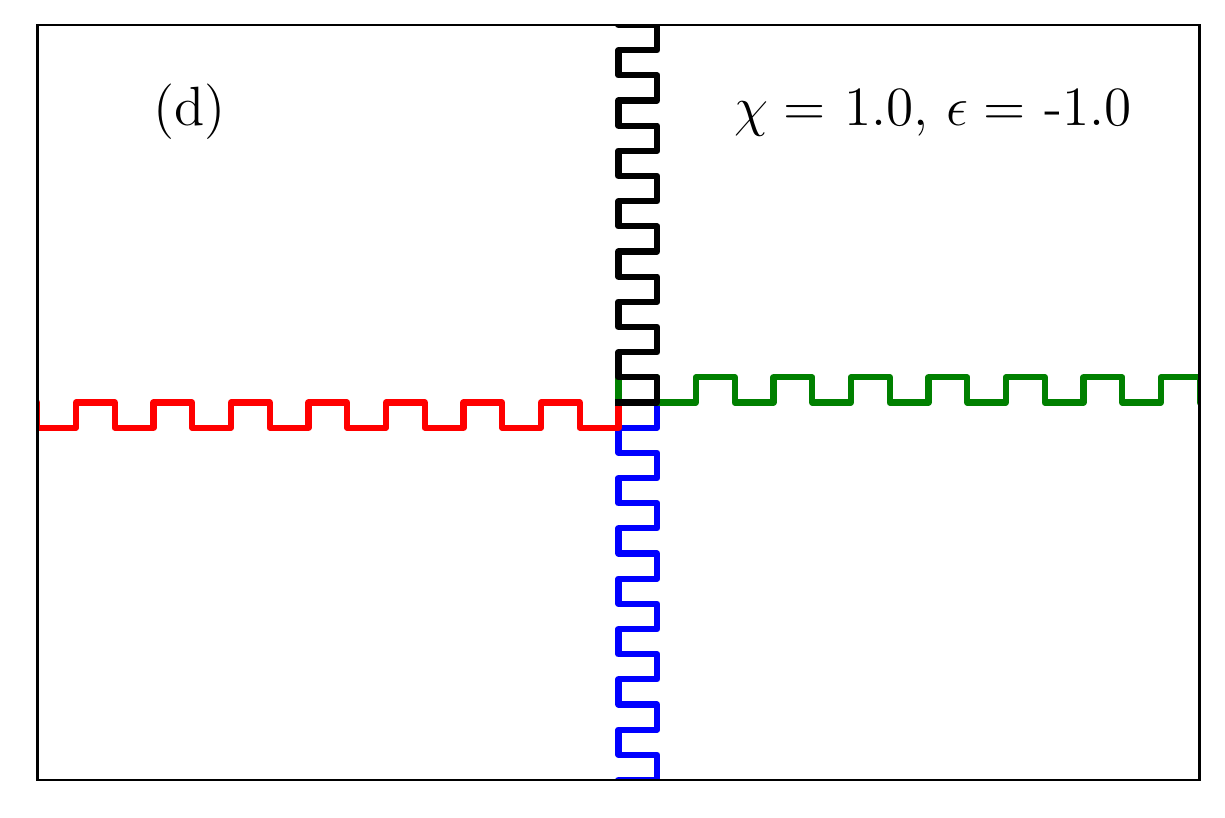}
    \caption{Typical trajectories with both interactions. Parameters: $N=500$ and $\beta = 100$. At each graph, there are four random trajectories indicated by the colors: blue, red, green and black. The values of $\chi$ and $\varepsilon$ are indicated on each graph. (a) $\chi = \varepsilon = 1.0$. (b) $\chi  = -1.0$ and $\varepsilon = 1.0$. (c) $\chi = \varepsilon = -1.0$. (b) $\chi = 1.0$ and $\varepsilon = -1.0$.}
    \label{figura6}
\end{figure*}

Figure \ref{figura6} shows typical trajectories considering the competition of both interactions. If $\chi =-1.0$ (Figs. \ref{figura6}(b) and (c)), the folding interaction prevails regardless the value of $\varepsilon$ and the trajectories are vertical and horizontal straight lines. On the other hand, if $\chi = 1.0$ ((Figs. \ref{figura6}(a) and (d)) the trajectories becomes completely filled with folding: no straight lines (although the general trend is a straight line). In this case, if $\varepsilon =1.0$ the trajectories make an angle of 45 degrees with the horizontal axes. This happens because the contact interaction avoids the contacts here. However if $\varepsilon = -1.0$ the contacts will be more favourable, which results in horizontal and vertical trajectories.

\begin{figure}
    \centering
    \includegraphics[scale = 0.62]{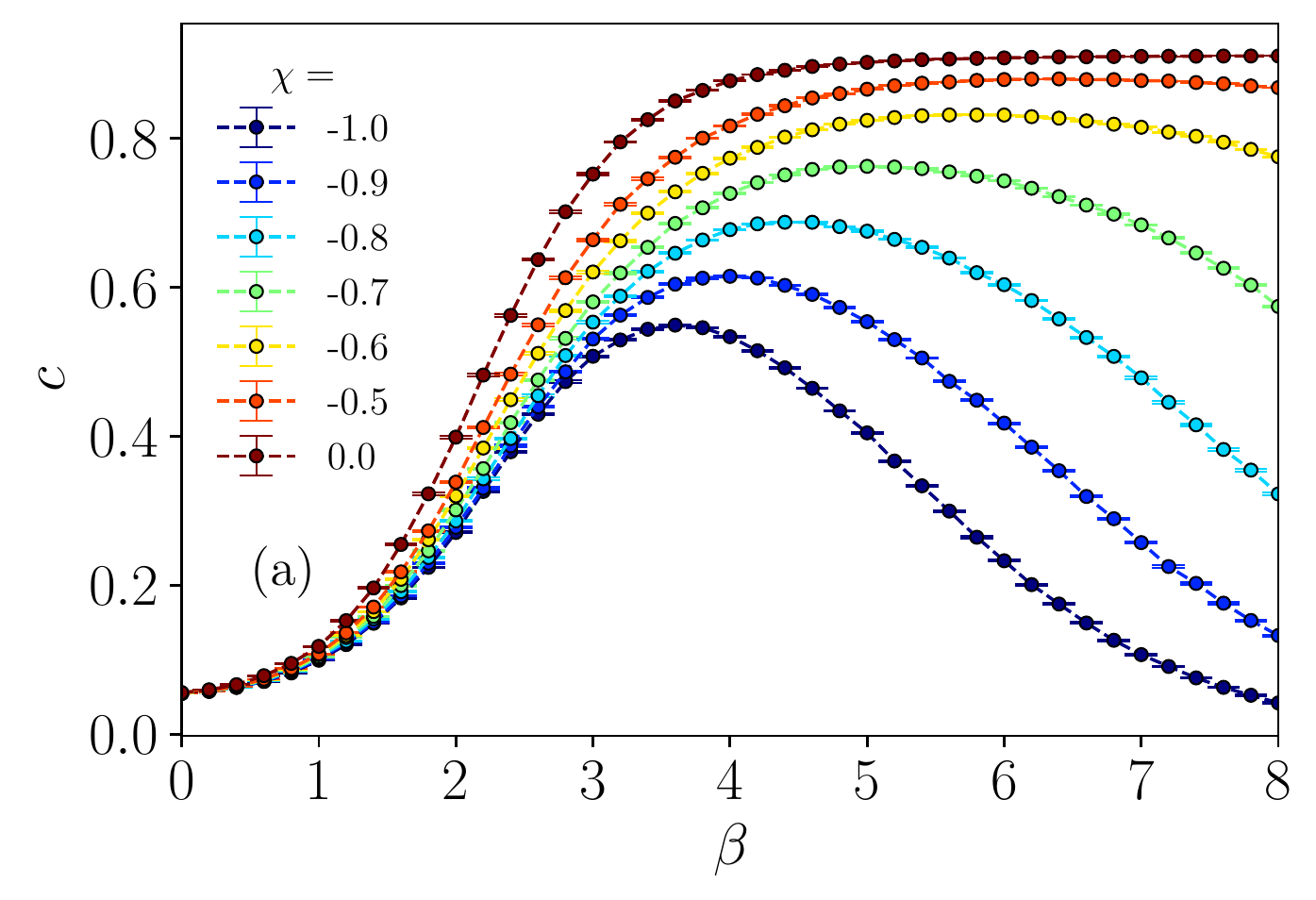}
    \includegraphics[scale = 0.62]{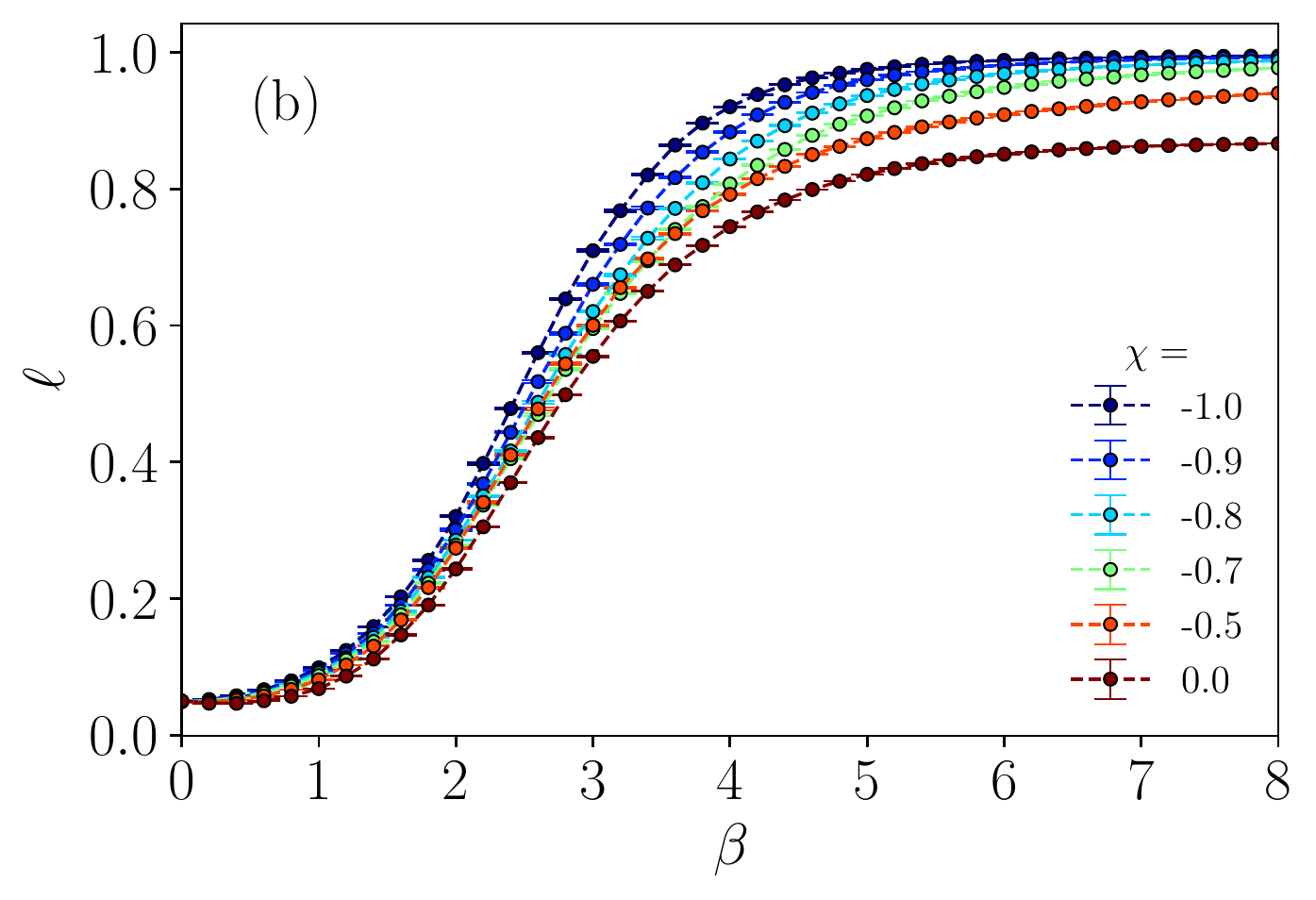}
    \caption{Behavior of the order parameters vs temperature $\beta$ as the folding interaction is increased. Parameters: $N=500$, $S_1 = 50$, $S_2=500$ and $\varepsilon = -1.0$ (contact interaction). (a) Number of contacts $c$. (b) Number of straight lines $\ell$.}
    \label{figura7a}
\end{figure}

As the results so far show that the folding interaction is dominant over the contact one, it is interesting to analyze how the system behaves as the former is gradually added to the system. Starting from the configuration with contact interaction only ($\varepsilon = -1.0$ and $\chi = 0.0$), we analyzed the system as the folding interaction is enhanced (increasing $\chi$). Fig. \ref{figura7a}(a) shows that the number of contacts $c$ is a smooth step shape considering only the contact interaction (dark red markers), which is expected. Ate $\beta > 4$, $c$ reaches its maximum value. However, as the folding interaction is considered, increasing $\chi$, the number of contacts at large $\beta$ decreases, as the folding one is dominant over the contact one. In the end, with $\chi = -1.0$, the phase for large $\beta$ is the extended one and $c$ tends to zero.

On the other hand, the number of straight lines $\ell$ shows a smaller change, as shown on Fig. \ref{figura7a}(b). The case $\chi = 0.0$ refers to a trajectory like the red one on Fig. \ref{figura2}(a), while the case $\chi = -1.0$ refers to the red one on Fig. \ref{figura2}(b). In both cases, clearly, there is a large number of straight lines, creating a large value of $\ell$ at large $\beta$.

\section{Conclusions}

We analyzed the self-avoid random walk algorithm with contact and folding interactions in order to study the effects of these interaction on the system. We observed that at low temperatures the contact interaction creates the compact phase and the folding one creates the extended phase. Each interaction has its own order parameter, defined after the typical event. The control parameter of each interaction defines if the typical event will be favoured or avoided in each case. When the two interactions are present, the folding one is dominant over the contact one. However, even though the trajectories are extended, the contacts are present. This works contributes to the basic understanding of the less studied folding interaction on the system, which is an important factor to the phase diagram of polymers.

\section{Acknowledgements}

PF Gomes and AA Costa acknowledges the support of CNPq (grant 405508/2021-2) and PF Gomes also acknowledges the support of FAPEG.

\appendix

\section{Appendix}

\section{Main algorithm}

In this section we present more details about the main algorithm which calculates the trajectory. Each trajectory is composed of $n$ pairs $(x_i,y_i)$ of points on the $xy$ plane, with $i=0,1,2,3,...,n-1$. These pairs are stored as two vectors $v_x$ and $v_y$. The input of the algorithm are the parameter $N$ and the interaction control parameters $\varepsilon$ and $\chi$.
The algorithm of the $F_1$ function is:
\begin{enumerate}
    \item Input parameters: $N,\varepsilon,\chi$.
    \item Node 0: position $v_x(0)=v_y(0)=0$.
    \item Node 1: calculate the possible positions $P$. All of them have the same probabilities. The node is randomly chosen and added to the trajectory $(v_x,v_y)$.
    \item Add these two nodes to the used points list $U$ that keeps the positions belonging to the trajectory.
    \item Calculate the possible positions $P$ for the next node.
    \item Considering the possible and the used positions, from the lists $P$ and $U$, calculate the available positions for the next node. The positions already on the trajectory are excluded from the list $A$.
    \item Calculate the number of contacts $c_i$ and straight lines $\ell_i$ for each available node (from the list $A$).
    \item Calculate the final probabilities $p_i$ for each available node and choose the next node.
    \item Update the trajectory $(v_x,v_y)$ and the used nodes list $U$.
    \item Perform the steps 5 to 9 $N-2$ times more.
    \item If the at any time the list $A$ is calculated empty, it means there is no available nodes and the trajectory is trapped. In this case, the algorithm must halt.
\end{enumerate}

The list $P$ and $A$ should be calculate at every step. The list $U$ and the vectors $(v_x,v_y)$ are created in the beginning and updated at every step.

\section{Arbitrary probabilities}

Once the set of probabilities $p_i$ are calculated, the new node must be chosen accordingly. If there is only one available node, it will be chosen. If there are two or three, the next one must be chosen using the probabilities. In this description we use $\Gamma$ as a representation of a random number between 0.0 and 1.0 generated from a uniform distribution. This is done in the following way:
\begin{enumerate}
    \item We generate a random number $0.0 < \Gamma < 1.0$. 
    \item If $\Gamma < p_1$ we choose the node 1. 
    \item If $p_1 < \Gamma < p_1+p_2$ we chose node 2. 
    \item If $p_1+p_2 < \Gamma < p_1+p_2+p_3$ we chose node 3. 
\end{enumerate}
If there are only two nodes, the step 4 is not performed.


\begin{thebibliography}{}

\bibitem{Flory1949} P. J. Flory. Jour. of Chem. Phys. \textbf{17}, 303 (1949). 

\bibitem{Flory1953} P. J. Flory \textit{Principles of Polymer Chemistry}, Cornell University Press, Ithaca, New York, (1953).

\bibitem{Canevarolo2002} S. V. Canevarolo Jr. \textit{Ci\^{e}ncia dos Pol\'{i}meros}. Editora Artliber, 2nd edition (2002).

\bibitem{Gedde2019} U. W. Gedde and M. S. Hedenqvist \href{https://doi.org/10.1007/978-3-030-29794-7}{\textit{Fundamental Polymer Science}}, Springer 2nd edition.

\bibitem{Micheletti2011} C. Micheletti, D. Marenduzzo, E. Orlandini.  Physics Reports \textbf{504}, 1-73 (2011).

\bibitem{Gennes1979} P.-G. de Gennes \textit{Scaling Concepts in Polymer Physics}. Cornell University Press (1979).

\bibitem{Oliveira2016} T. J. Oliveira and J. F. Stilck.  Phys. Rev. E \textbf{93}, 012502 (2016).

\bibitem{Duplantier1987} B. Duplantier and H. Saleur. Phys. Rev. Let. \textbf{59}, 539 (1987).

\bibitem{Grassberger1995} P. Grassberger and R. Hegger. Jour. Chem. Phys. \textbf{102}, 6881 (1995).

\bibitem{Grassberger1997} P. Grassberger. Phys. Rev. E \textbf{56}, 3682 (1997).

\bibitem{Rubin1965} R. J. Rubin. The Journal of Chemical Physics, 43(7), 2392-2407 (1965).

\bibitem{Kholodenko1984} A. L. Kholodenko and K. F. Freed. J. Chern. Phys. \textbf{80}(2), 900 (1984).

\bibitem{Maes1990} D. Maes and C. Vanderzande. Phys. Rev. A  \textbf{41} 3074 (1990).

\bibitem{Vanderzande1998} C. Vanderzande \textit{Lattice Models of Polymers}, Cambridge University Press, Nova York (1998).

\bibitem{Foster2001} D. P. Foster and F. Seno. J. Phys. A: Math. Gen. 34 9939 (2001).

\bibitem{Micheletti2021} C. Micheletti, P. Hauke, and P. Faccioli, Phys. Rev. Let. \textbf{127}, 080501 (2021).

\bibitem{Vilela2020} E. B. Vilela, H. A. Fernandes, F. L. P. Costa, P. F. Gomes, Journal of Computational Chemistry \textbf{41}, 1964–1972 (2020).

\bibitem{Fernandes2016} H. A. Fernandes, R. da Silva, E. D. Santos, P. F. Gomes, E. Arashiro. Phys. Rev. E \textbf{94} (2016) 022129.

\bibitem{Chowdhury1985} D. Chowdhury and B. K. Chakrabarti Journal of Physics A: Mathematical and General \textbf{18} L377 (1985).


\bibitem{Rockenbach2010} R. Rockenbach and R. A. Zara. Rev. Bras. de Ens. de F\'{i}s. \textbf{32}, 4305 (2010).

\bibitem{Clisby2010} N. Clisby. Phys. Rev. Lett. \textbf{104}, 055702 (2010).

\bibitem{Jensen2004}  I. Jensen. J. Phys. A: Math. Gen. \textbf{37} 5503 (2004).

\bibitem{Krawczyk2010} J. Krawczyk, A. L. Owczarek, T. Prellberg. A semi-flexible attracting segment model of two-dimensional polymer collapse. Physica A \textbf{389} 1619–1624 (2010).

\bibitem{Madras2013} N. Madras and G. Slade. The Self-Avoiding Walk. Birkh\"{a}user (2013).

\bibitem{Montroll1950} E. W. Montroll. The Journal of Chemical Physics \textbf{18}, 734-743 (1950).


\bibitem{Narasimhan2001} S. L. Narasimhan, P. S. R. Krishna, K. P. N. Murthy, and M. Ramanadham. Phys. Rev. E, \textit{65}, 010801 (2001).


\bibitem{Taylor1998} M. P. Taylor and J. E. G. Lipson. Journal of Chemical Physics \textbf{109} 7583 (1998).


\bibitem{Krawczyk2009} J. Krawczyk, A. L. Owczarek, T. Prellberg.  Physica A \textbf{388} (2009) 104–112 .


\bibitem{Rockenbach2014} R. Rockenbach and R. A. Zara. Revista Brasileira de Ensino de F\'{i}sica \textbf{36}, 4307 (2014).


\bibitem{Bastolla1997} U. Bastolla and P. Grassberger. Journal of Statistical Physics, Vol. \textbf{89}, (1997) 1061.

\bibitem{Amit1983} D. J. Amit, G. Parisi, and L. Peliti. 
Phys. Rev. B \textbf{27}, (1983) 1635 .

\bibitem{Grassberger2017} P. Grassberger. Physical Review Letters \textbf{119}, 140601 (2017).

\bibitem{Harris2020} C. R. Harris, \textit{et al},  Array programming with Numpy, Nature \href{https://doi.org/10.1038/s41586-020-2649-2}{\textbf{585}, 357–362 (2020)}.

\bibitem{Hunter2007} J. D. Hunter. Matplotlib: A 2d graphics environment, Computing in Science \& Engineering \href{https://doi.org/10.1109/MCSE.2007.55}{\textbf{9}, 90-95 (2007)}.

\bibitem{McKinney2010} W. McKinney. Data structures for statistical computing in python, in: Proceedings of the 9th Python in Science Conference \href{http://dx.doi.org/https://doi.org/10.25080/Majora-92bf1922-00a}{\textbf{445}, 51-56 (2010)}.
\end{thebibliography}
\end{document}